\documentclass[12pt, twocolumn]{aastex63}
\usepackage{physics}
\usepackage[normalem]{ulem}
\usepackage{amsmath, amssymb}
\usepackage{bbold}
\definecolor{beaver}{rgb}{0.62,0.51,0.44}
\newcommand{\LCDM}{\ensuremath{\Lambda\mathrm{CDM}}}

\shorttitle{Model-independent constraints on Type Ia supernova light-curve hyper-parameters}
\shortauthors{Koo, Shafieloo, Keeley \& L'Huillier}

\begin{document}

\title{Model-independent constraints on Type Ia supernova light-curve hyper-parameters and reconstructions of the expansion history of the Universe}

\correspondingauthor{Arman Shafieloo}
\email{hkoo@kasi.re.kr, shafieloo@kasi.re.kr, rkeeley@kasi.re.kr, blhuillier@yonsei.ac.kr}

\author[0000-0003-0268-4488]{Hanwool Koo}
\affil{Korea Astronomy and Space Science Institute, Daejeon 34055, Korea}
\affil{University of Science and Technology, Yuseong-gu 217 Gajeong-ro, Daejeon 34113, Korea}

\author[0000-0001-6815-0337]{Arman Shafieloo}
\affil{Korea Astronomy and Space Science Institute, Daejeon 34055, Korea}
\affil{University of Science and Technology, Yuseong-gu 217 Gajeong-ro, Daejeon 34113, Korea}

\author[0000-0002-0862-8789]{Ryan E. Keeley}
\affil{Korea Astronomy and Space Science Institute, Daejeon 34055, Korea}

\author[0000-0003-2934-6243]{Benjamin L'Huillier}
\affil{Yonsei University, Seoul 03722, Korea}

\date{\today}

\begin{abstract}

We reconstruct the expansion history of the Universe using type Ia supernovae (SN Ia) in a manner independent of any cosmological model assumptions.  To do so, we implement a non-parametric iterative smoothing method on the Joint Light-curve Analysis (JLA) data while exploring the SN Ia light-curve hyper-parameter space by Monte Carlo Markov Chain sampling.
We test to see how the posteriors of these hyper-parameters depend on cosmology, whether using different dark energy models or reconstructions shift these posteriors.
Our constraints on the SN Ia light-curve hyper-parameters from our model-independent analysis are very consistent with the constraints from using different parametrizations of the equation of state of dark energy, namely the flat $\LCDM$ cosmology, the Chevallier-Polarski-Linder (CPL) model, and the Phenomenologically Emergent Dark Energy (PEDE) model. This implies that the distance moduli constructed from the JLA data are mostly independent of the cosmological models. We also studied that the possibility the light-curve parameters evolve with redshift and our results show consistency with no evolution. 
The reconstructed expansion history of the Universe and dark energy properties also seem to be in good agreement with the expectations of the standard $\LCDM$ model. However, our results also indicate that the data still allow for considerable flexibility in the expansion history of the Universe.

\end{abstract}


\section{Introduction}\label{sec:intro}

The concordance model of cosmology, $\LCDM$ ($\Lambda$ for the cosmological constant and CDM for the cold dark matter), is based on general relativity and the assumption that the Universe is isotropic and homogeneous.
It has successfully explained various astronomical observations at least from the epoch of Big Bang nucleosynthesis. 
This model explains the low-redshift dynamics of the Universe with only two parameters, the Hubble constant, $H_0$, and the matter density, $\Omega_{\rm m}$.
Despite the simplicity of the model, most astronomical observations are in good agreement with the concordance model and so far there has been no strong observational evidence against it.

SN Ia distance measurements \citep{riess-etal98,perlmutter-etal99} have become one of the most important datasets of modern cosmology since they are standardizable candles and they 
directly measure the accelerating expansion of the Universe at late times. Previous analyses using SN Ia compilations including SuperNova Legacy Survey  \citep[SNLS,][]{sullivan-snlsc05}, Gold \citep{riess-etal07}, Union \citep{kowalski-etal08}, Constitution \citep{hicken-etal09}, Union2 \citep{amanullah-etal10}, Union2.1 \citep{suzuki-etal12}, Joint Light-curve Analysis \citep[JLA,][]{betoule-etal14} and Pantheon \citep{scolnic-etal18}, have been mostly compatible with the flat $\LCDM$ model\footnote{We should note that a few previous studies, including \cite{NielsenSarkar,kim-etal18,kim-etal19}, find that deviations from $\Lambda$CDM or a non-accelerating universe can be still allowed by SN Ia data.}. Most of these analyses are based on the SALT \citep{guy-etal05} or SALT2 \citep{guy-etal07,mosher-etal14} SN Ia light curve models, which are
compressions of the full light-curves and are trained on simulated SN Ia datasets.
It outputs three parameters that summarize the information contained in the SN Ia light-curves, $m_B$, the B-band brightness, $X_1$, the stretch factor, and $C$, the color of the SN.

Though the nature of dark energy still remains elusive, analyses of SN Ia typically assume some sort of model for the dark energy component, and in most cases this model is the concordance spatially flat $\Lambda$CDM model.
It is possible that the assumption of $\LCDM$ is biasing the light-curve hyper-parameters towards certain values and also biasing inferences of the expansion history of the Universe, since the standardization of SN Ia is purely empirical. 
Thus, we seek to relax the assumption of $\LCDM$, or any parametric form of the expansion history,  and explore what values of the light-curve hyper-parameters might result.  This, in turn, allows us to explore possibly different expansion histories of the Universe.
Reconstructions of the distance moduli and expansion history of the Universe can, in fact, be carried out purely kinematically by using a non-parametric method without assuming a particular cosmological model or gravitational model.

The main goal of this paper is to provide constraints on SN Ia light-curve hyper-parameters without any cosmology model assumption followed by reconstruction of the expansion history of the universe and the properties of dark energy. We use a non-parametric iterative smoothing method first introduced by \cite{shafieloo-etal06,shafieloo07} and improved further by \cite{shafieloo-clarkson10,shafieloo-etal18} to reconstruct the distance moduli, expansion history of the Universe, and dark energy properties. We choose to use the JLA compilation rather than the more recent Pantheon \citep{scolnic-etal18} because JLA provides the shape and color information of the light curve, which are essential to constrain the light-curve hyper-parameters. In Sec.~\ref{sec:corner} we describe the method we use to constrain the light-curve hyper-parameters, namely $\alpha, \beta, M_B$, and $\Delta_M$ \citep{betoule-etal14}. Reconstructions of the luminosity distance, expansion history of the Universe, $Om$ parameter, and deceleration parameter are shown in in Sec.~\ref{sec:recon}, and our discussions and conclusions are presented in Sec.~\ref{sec:dis}.

\section{Cosmological constraints on light-curve hyper-parameters}\label{sec:corner}

The first step in a cosmological analysis of SN Ia is to standardize them. Using the SALT2 light-curve parameters from the JLA compilation, we calculate the distance moduli ($\mu$) as in \cite{betoule-etal14}, by using the standard Tripp formula \citep{Tripp98}, which is a linear scaling of the light-curve parameters,
\begin{equation}\label{eqn:light_curve}
\mu=m_B^\star-(M_B-\alpha X_1+\beta C).
\end{equation}
For each SN, the SALT2 fitter yields three light-curve parameters, the observed B-band peak magnitude ($m_B^\star$), the stretching of the light curve ($X_1$), and the SN color at maximum brightness ($C$). $M_B$, $\alpha$, and $\beta$ are hyper-parameters that need to be marginalized over when making inferences about cosmology. Following \cite{betoule-etal14}, we account for the absolute magnitude's ($M_B$) dependence on host galaxy properties with the following prescription, first proposed by \cite{Conley-11}

\begin{equation}\label{eqn:M_B}
  M_B = \left\{\begin{array}{lr}
        M_B^1, & \text{if } M_{\mathrm{stellar}}<10^{10}M_{\mathrm{sun}}\\
        M_B^1 +\Delta_M, & \text{otherwise } 
        \end{array}\right\} .
\end{equation}
With a prescription for how to map the light-curve parameters to distance moduli in hand, we can now investigate what these distances have to say about cosmology and what potential correlations between light-curve modeling and cosmology there might be. 

In a flat FLRW Universe with a dark energy component with equation-of-state parameter $w(z)$, the luminosity distance can be written as

\begin{equation}\label{eqn:d_l_model}
d_L(z) = \frac{c}{H_0}(1+z)\int_{0}^{z}\frac{dz'}{h(z')}
\end{equation}
where the expansion history is
\begin{equation}\label{eqn:h_z_model}
h^2(z) = \Omega_m(1+z)^3 + (1-\Omega_m)\exp(3\int_{0}^{z}\frac{1+w(z')}{1+z'}dz').
\end{equation}
In \LCDM, $w=-1$, but in general it can vary. For instance, CPL~\citep{chevallier-polarski01,linder03} is a parameterization of the evolution of $w(z)$ near $z=0$ where $w(z)=w_0+w_a\tfrac{z}{1+z}$.  
The PEDE model, recently introduced by \cite{li-shafieloo19}, offers another example about how evolving equation of state, 
$w(z)=\,-\tfrac{1}{3\ln 10} \left({1+\tanh\left[\log_{10}\,(1+z)\right]}\right)-1$.  We train each of these models on the JLA data and test to see if the inferred light-curve hyper-parameters differs between these models.  Since $M_B$ is exactly degenerate with $H_0$, SN Ia alone cannot constrain an absolute distance scale. Therefore, without loss of generality, we fixed $H_0=70\ \rm km\ s^{-1}\ Mpc^{-1}$ for all of these models.  Should we choose a different value for fixing $H_0$, then the posterior of $M_B$ would shift as $M_B + 5\log_{10}(H_0/70)$.

In addition to these phenomenological models, we also look at how the light-curve hyper-parameters behave when the cosmological distances are calculated using a model-independent reconstruction method.
Namely, we apply the iterative smoothing method \citep{shafieloo-etal06,shafieloo07,shafieloo-clarkson10,shafieloo-etal18} to the distance moduli derived from the JLA data. The iterative smoothing method is a model-independent approach to reconstruct the distance moduli from the SN Ia data.

We start from some initial guess $\hat{\mu}_0$, and derive the reconstructed distance moduli $\hat{\mu}_{n+1}$ repeatedly at iteration $n+1$ as

\begin{equation}\label{eqn:smooth}
\hat{\mu}_{n+1}(z) = \hat{\mu}_n(z) + \frac{\boldsymbol{\delta\mu_n}^T \cdot \mathbf{C^{-1}} \cdot \boldsymbol{W}(z)}{\mathbb{1}^T \cdot \mathbf{C^{-1}} \cdot \boldsymbol{W}(z)}
\end{equation}
where $\mathbb{1}^T=(1,\cdots,1)$, the weight $\boldsymbol{W}$ and residual $\boldsymbol{\delta\mu_n}$ denote

\begin{equation}\label{eqn:weight}
\boldsymbol{W}_i(z)=\exp(-\frac{\ln^2(\frac{1+z}{1+z_i})}{2\Delta^2})
\end{equation}

\begin{equation}\label{eqn:residual}
\boldsymbol{\delta\mu_n}|_i = \mu_i -  \hat{\mu}_n(z_i)
\end{equation}
and $\mathbf{C^{-1}}$ indicates the inverse of the covariance matrix of the JLA data. The smoothing width is set to $\Delta = 0.3$ for the analysis of the JLA data following \cite{shafieloo-etal06, 2017JCAP...01..015L, lhuillier-etal18}. The covariance matrix includes not only the statistical light-curve fit uncertainty but the systematic uncertainties from the calibration, light-curve model, bias correction, mass step, and extinction by our galaxy. It also includes systematic uncertainties caused by peculiar velocities, gravitational lensing, and unknown sources.

We define the $\chi^2$ value of the reconstruction $\hat{\mu}_n(z)$ as

\begin{equation}\label{eqn:chi2}
\chi_n^2 = \boldsymbol{\delta\mu_n}^T \cdot \mathbf{C^{-1}} \cdot \boldsymbol{\delta\mu_n},
\end{equation}
and each iteration will, by construction, produce a reconstruction with a better $\chi^2$ than the previous iteration.

\begin{figure}
\centering
\includegraphics[width=0.45\textwidth]{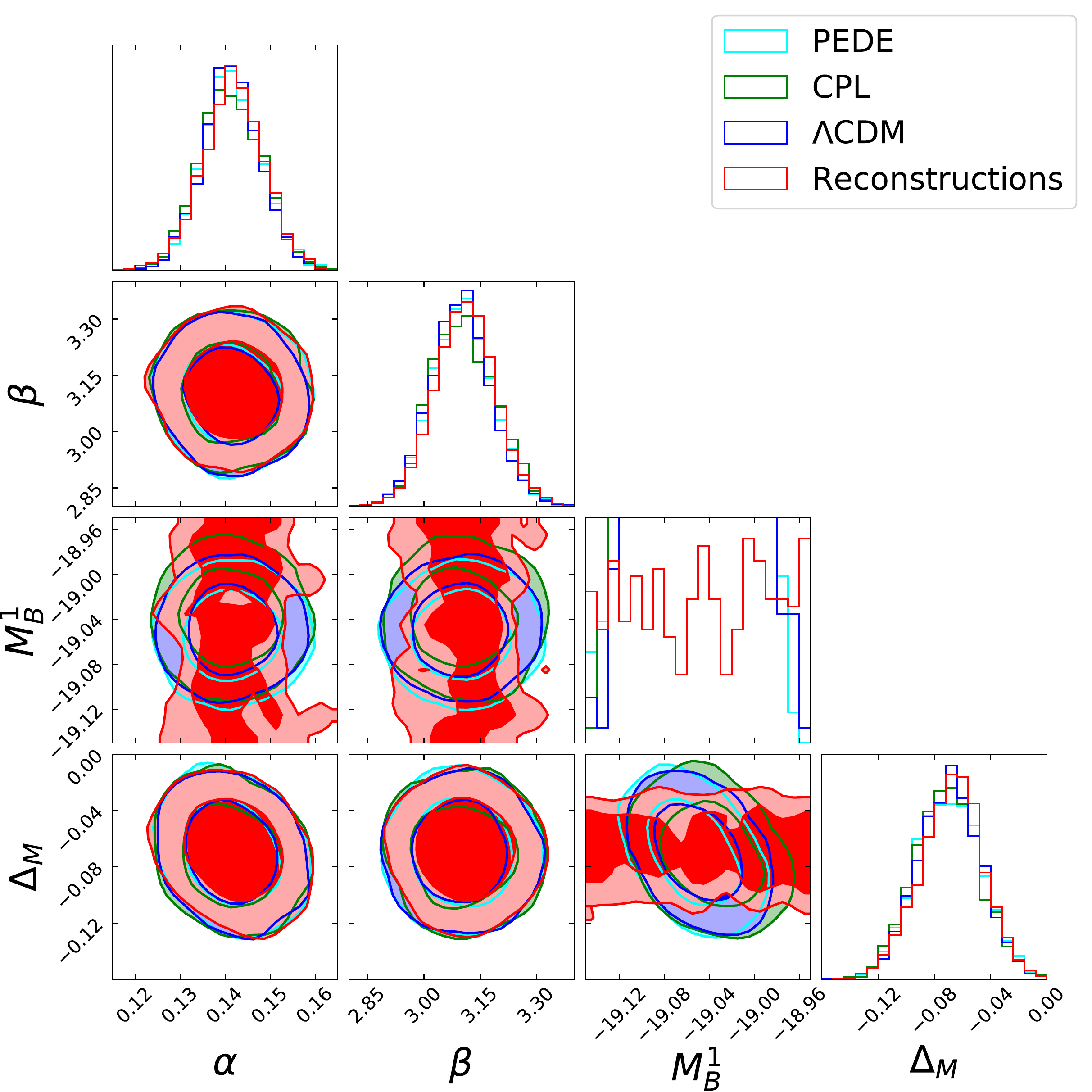}
\caption{\label{fig:contours}Cosmological constraints on the light-curve hyper-parameters from the iterative smoothing method (red), flat $\LCDM$ model (blue), CPL model (green), and PEDE model (cyan), using the JLA data. Note that $M_B^1$ is not constrained for the iterative smoothing case since for any change in $M_B^1$ the iterative smoothing method can shift $\hat{\mu}_n(z_i)$ by some constant value such that the $\chi^2$ value is unchanged.}
\end{figure}

We then use \texttt{emcee}~\citep{emcee} to perform a Markov chain Monte Carlo (MCMC) analysis in the light-curve hyper-parameter space and marginalize over the cosmological parameters for each of the parametric models as well as for the results from the iterative smoothing method. The resulting posteriors are shown in Fig.~\ref{fig:contours}.  
These light-curve hyper-parameters posteriors show no significant differences between the different cosmological models 
except that $M_B^1$ is not constrained for the iterative smoothing case since for any change in $M_B^1$ the iterative smoothing method can shift $\hat{\mu}_n(z_i)$ by some constant value such that the $\chi^2$ value is unchanged.
This lends credence to the idea that, for the JLA compilation, the light-curve hyper-parameters, and hence the distances derived from them, are independent of cosmology.

\begin{figure}
\centering
\includegraphics[width=0.45\textwidth]{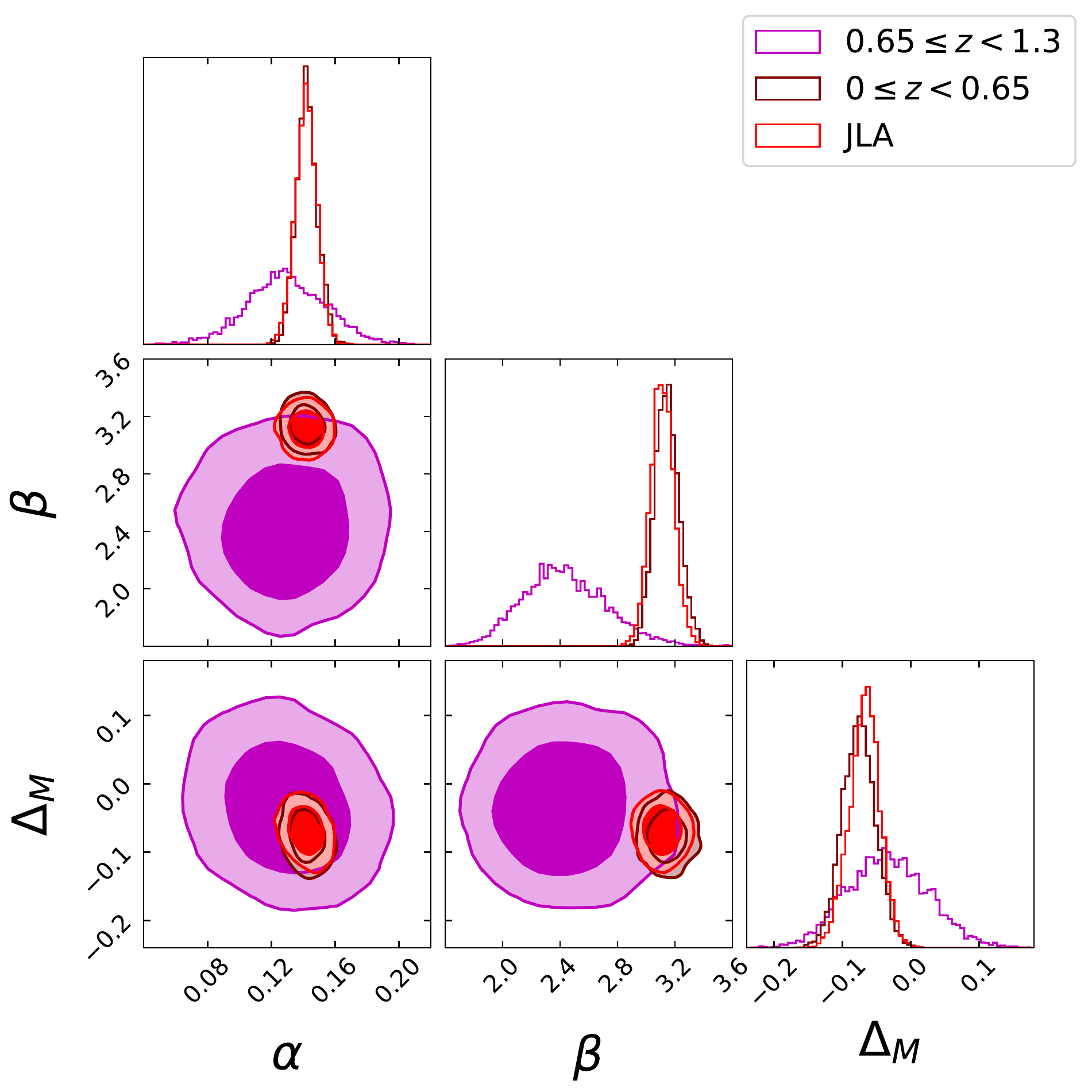}
\caption{\label{fig:contours_zb}Cosmological constraints on the light-curve hyper-parameters from the iterative smoothing method using JLA data with $0 \leq z < 0.65$ (maroon), with $0.65 \leq z < 1.3$ (magenta), and with all of them. We can see that there is no statistically meaningful redshift evolution in the light-curve hyper-parameters and the constraints are mainly influenced by the data in the low redshift bin.}
\end{figure}

Next, in order to see the possible evolution of the light-curve hyper-parameters, we split the full JLA compilation into two bins of equal width in redshift, $0 \leq z < 0.65$ and $0.65 \leq z < 1.3$. We then used our iterative smoothing method to calculate the posteriors of the light-curve hyper-parameters for each of these two bins.  Should the posteriors of these two bins be different, that would serve as some evidence that the simple linear scaling of the SALT2 light-curve parameters is insufficient to extract the true cosmological distances. Fig.~\ref{fig:contours_zb} shows that the posteriors of these different subsamples are different, though not significantly so.
The lack of statistical deviation between the contours from the two redshift bins and the relative size of the contours implies that there is no meaningful redshift evolution in the light-curve hyper-parameters and that they are constrained mainly by the data included in the low redshift bin.

\section{Model-independent Reconstructions}\label{sec:recon}

Beyond being appropriately agnostic about model assumptions, the iterative smoothing method can be used to reconstruct cosmological quantities from the JLA data, such as the expansion history or deceleration parameter.  By construction, these quantities will deviate from $\Lambda$CDM as the iterative smoothing method will always return distance moduli that have better $\chi^2$ than the best-fit $\Lambda$CDM ones (as we can start the procedure with the best-fit $\Lambda$CDM as the initial guess). This allows us to explore what deviations from the $\Lambda$CDM expansion history are still allowed by the JLA data.

\begin{figure}
\centering
\includegraphics[width=0.45\textwidth]{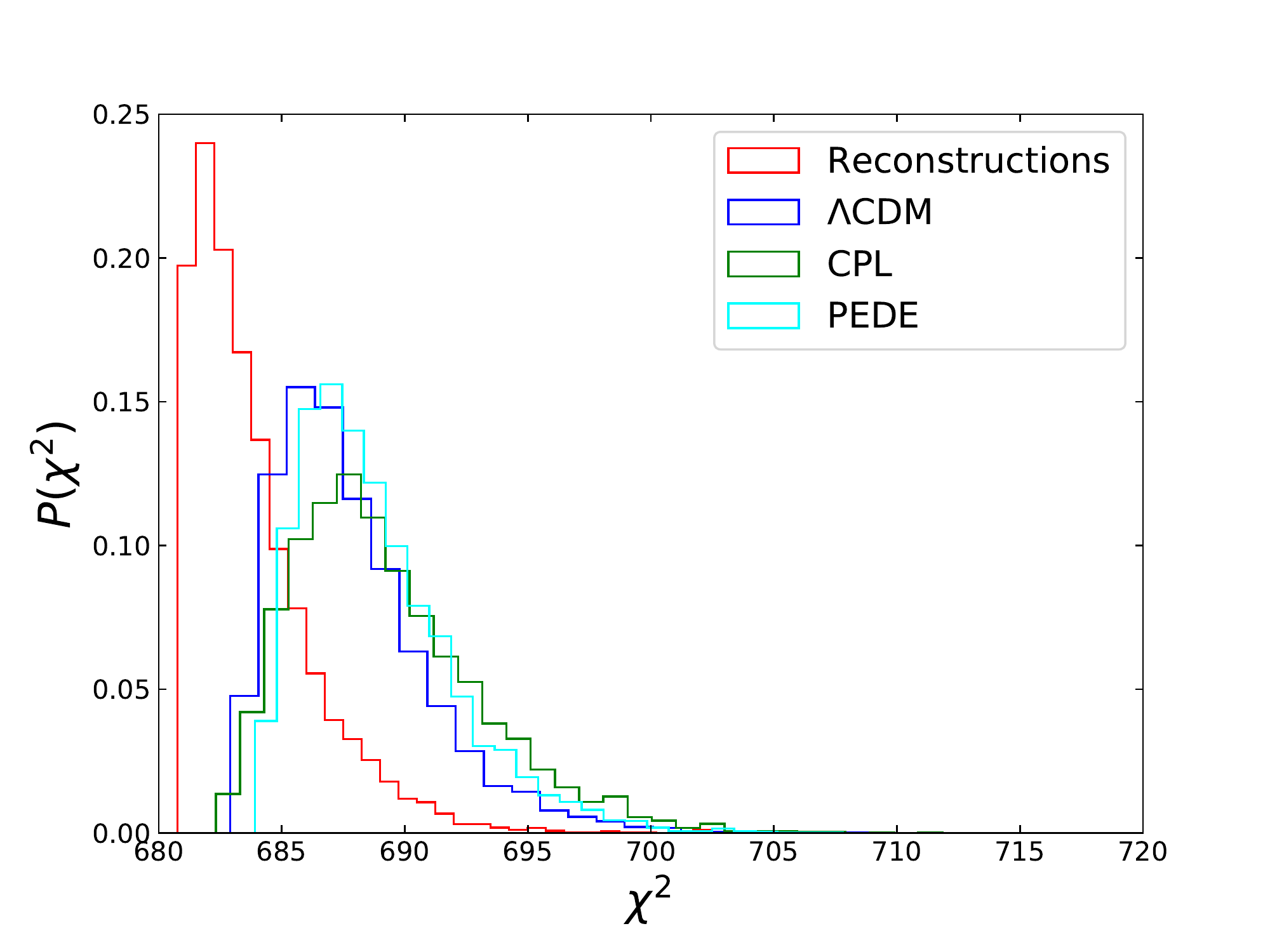}
\caption{\label{fig:pdf_chi2}Distribution of $\chi^2$ values for the iterative smoothing method (red), flat $\LCDM$ model (blue), CPL model (green), and PEDE model (cyan). These distributions are derived from the MCMC analysis allowing the light-curve hyper-parameters to vary.}
\end{figure}

Fig.~\ref{fig:pdf_chi2} shows the distribution of $\chi^2$ values for the three different models we considered and for the iterative smoothing method. These $\chi^2$ values are sampled from the posteriors of each case and so differences in these distributions can potentially be used as a metric for model selection. We will discuss about the issue of model selection in a separate future analysis. 
Further, this figure implies that most of the reconstructions from the iterative smoothing method have higher likelihoods than that of the predictions from $\LCDM$, which indicates that the data allows for some deviations from the standard $\Lambda$CDM model. 
This also shows the magnitude by which our beyond-$\Lambda$CDM reconstructions fit the data better.

From the reconstructed distance moduli $\hat{\mu}_n$, we can reconstruct 
the luminosity distance as $d_L(z) = 10^{\hat{\mu}_n\slash5-5}$ Mpc and
the expansion history as
 
\begin{equation}\label{eqn:exp_hist}
h(z) = \frac{c}{H_0}\left[\dv{}{z}\frac{d_L(z)}{1+z}\right]^{-1}.
\end{equation}

Next, from the reconstructed expansion history, we can reconstruct the $Om$ parameter, which is suggested by \citet{sahni-etal08} as a diagnostic for $\Lambda$CDM. Should $\Lambda$CDM be the true cosmology, this function would be constant in redshift and equal to $Om(z)=\Omega_m$. It is defined as

\begin{equation}\label{eqn:om_param}
Om(z) = \frac{h(z)^2-1}{(1+z)^3-1}.
\end{equation}
We can also reconstruct the deceleration parameter given by

\begin{equation}\label{eqn:decel_param}
q(z) = (1+z)\frac{\dv{h}{z}}{h} - 1.
\end{equation}

To explore the expansion history beyond the concordance model, we select 100 random light-curve hyper-parameter values from the MCMC chains and produced 80 reconstructions from each of these hyper-parameter combinations with different initial guesses and iterations of the smoothing. 
It has been demonstrated that the iterative smoothing method converges to the same smoothed function regardless of the initial guess or width of smoothing after enough number of iterations~\citep{shafieloo-etal06,shafieloo07,shafieloo-clarkson10}.
However, in the process of this convergence, many different forms of reconstructions can be generated, yet these reconstructions can still have a better likelihood than the best fit $\Lambda$CDM model. 
Hence starting from different initial guess models can reveal additional reconstructions that are both plausible and intriguing.
We perform the reconstruction for four different initial guess models. The models are: the best-fit flat $\LCDM$ model from the JLA, the flat sCDM model ($\Omega_m=1$), the flat $\LCDM$ model with $\Omega_m=0.01$, and the open empty Universe ($\Omega_k=1$). Aside from the best-fit $\Lambda$CDM, these models were chosen because they are meaningfully different from the best-fit $\Lambda$CDM.

The first few reconstructions from the alternative initial guess models do obviously have a worse likelihood than the best-fit $\Lambda$CDM, but 
as the number of iterations increases, the reconstructions would have a better likelihood than the best-fit $\Lambda$CDM. We emphasise again that regardless of the initial guess model, the smoothing procedure converges to the same reconstruction as the number of iterations increases \citep{shafieloo-etal06,shafieloo07,shafieloo-clarkson10}.

\begin{figure*}
\centering
\includegraphics[width=0.45\textwidth]{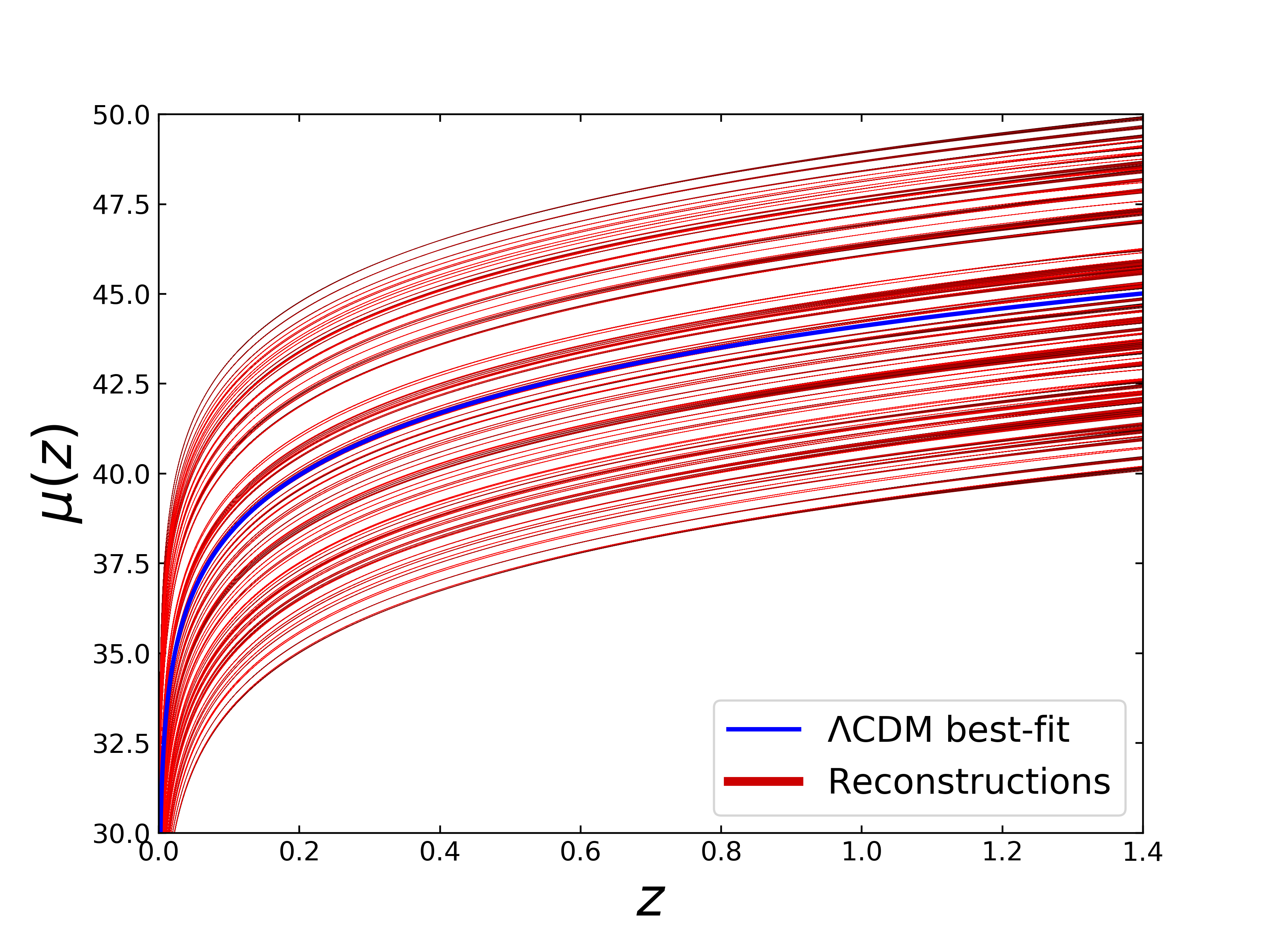}
\includegraphics[width=0.45\textwidth]{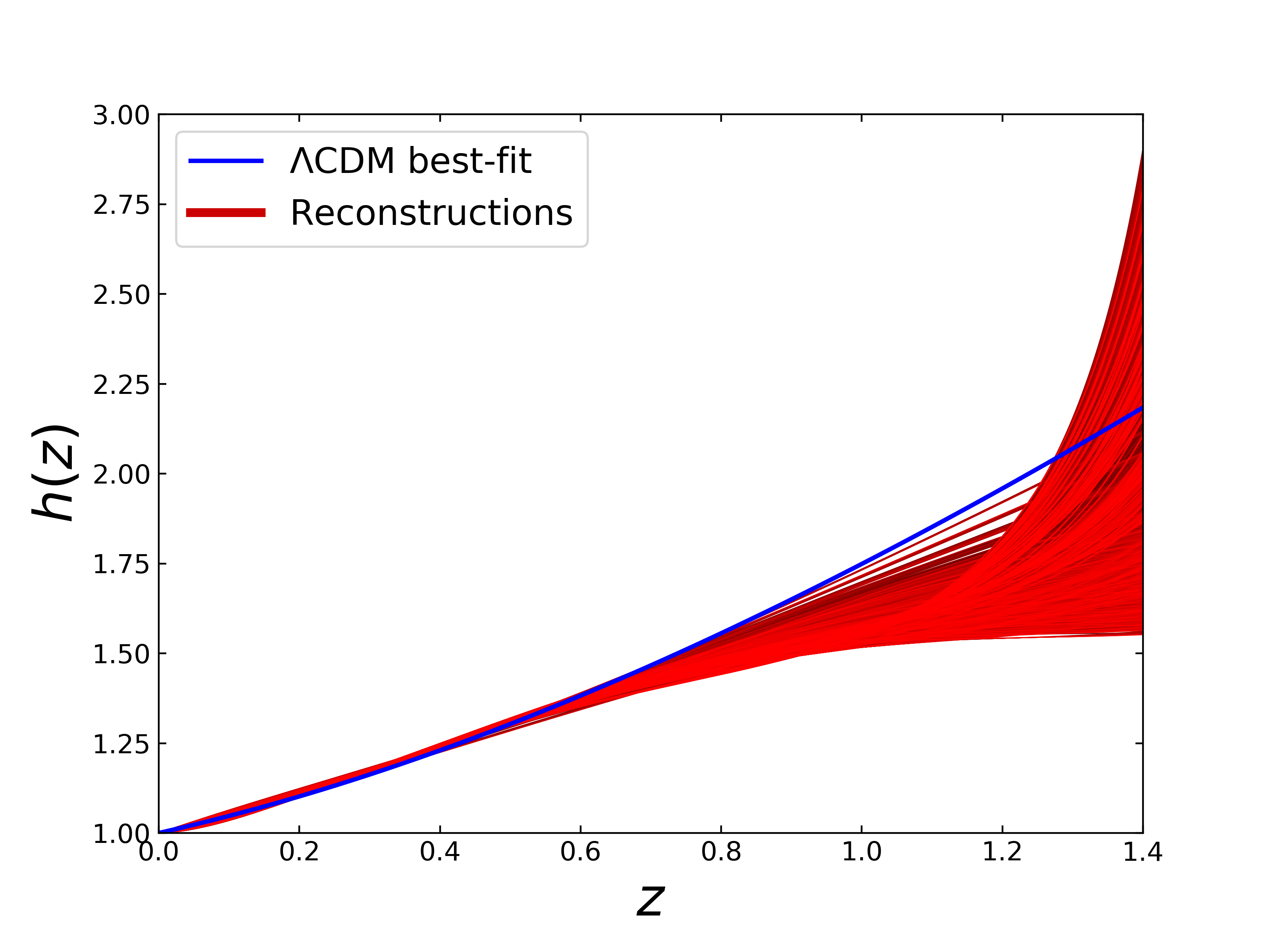}
\includegraphics[width=0.45\textwidth]{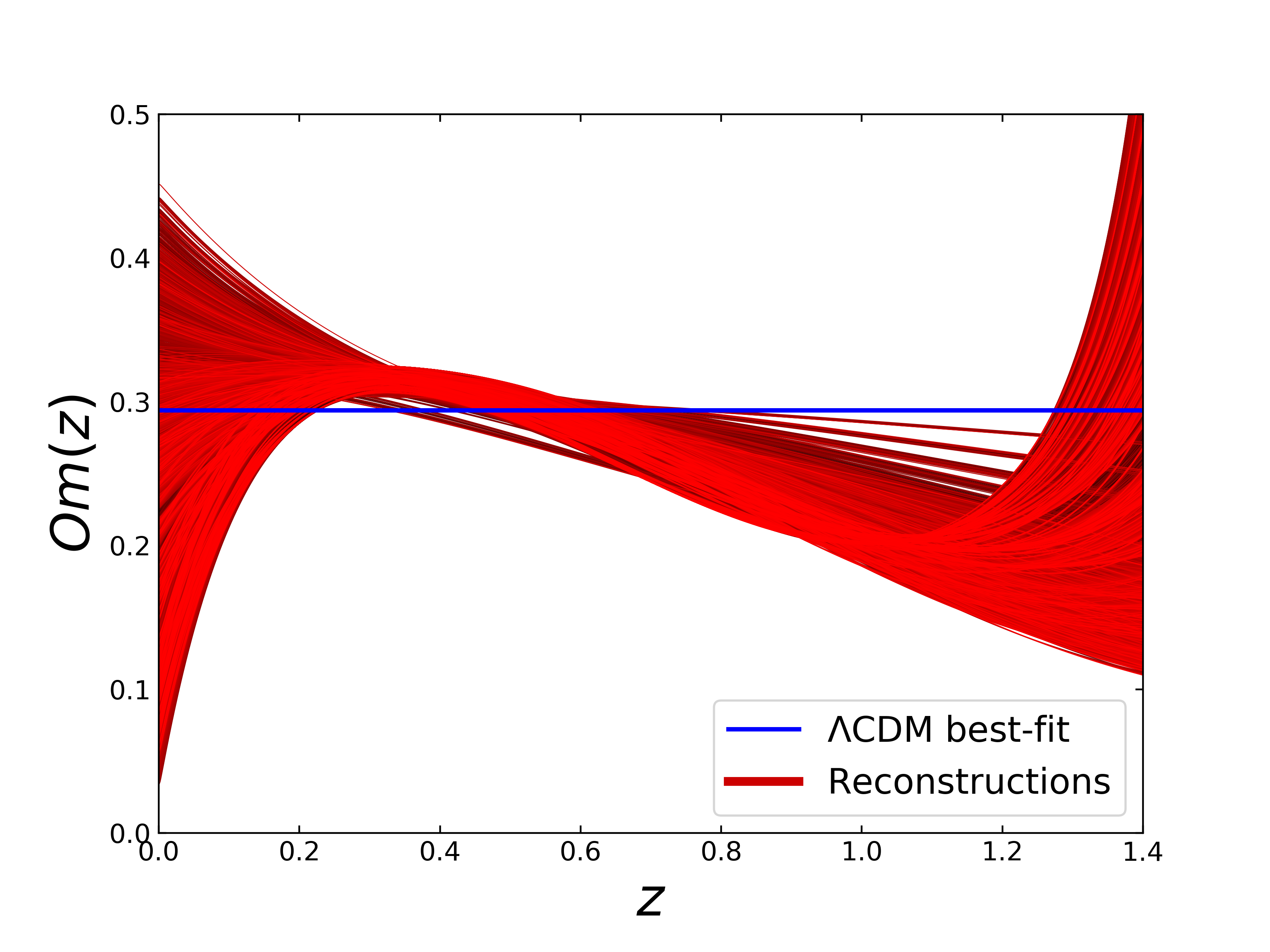}
\includegraphics[width=0.45\textwidth]{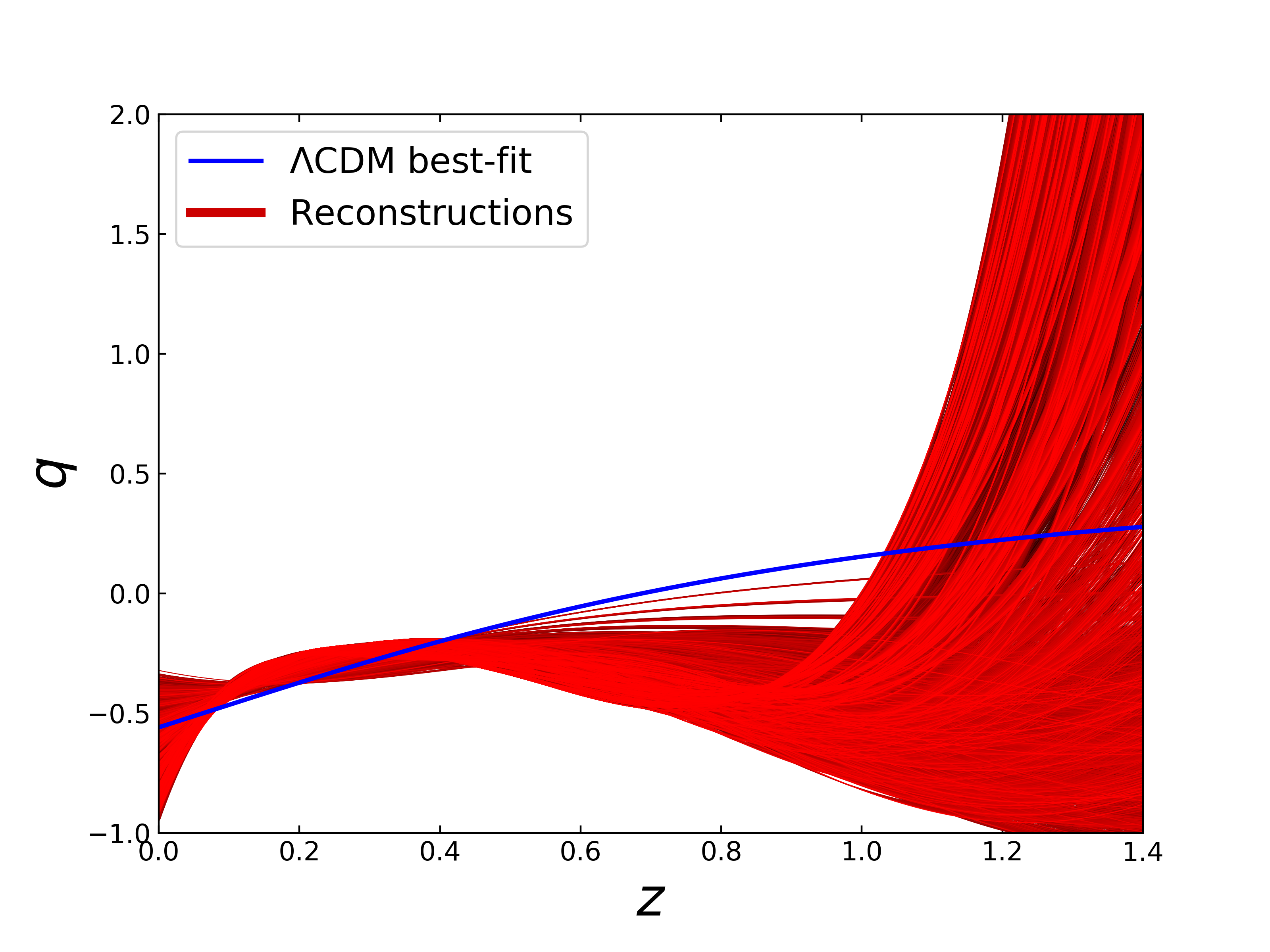}
\caption{\label{fig:recon}Reconstructions of the distance modulus $\mu(z)$, expansion history $h(z)$, Om parameter $Om(z)$, and deceleration parameter $q(z)$ using the iterative smoothing method on the JLA supernova data with $\chi^{2}<\chi^{2}_{\LCDM \rm{best-fit}}$. We can see that the data allows intriguing deviations from the $\Lambda$CDM model. The light-curve hyper-parameters are left free so the distance modulus displays an additive freedom.}
\end{figure*}

Fig.~\ref{fig:recon} shows 8000 reconstructions starting from the four different initial guesses that have higher likelihoods than the best-fit $\LCDM$ model. 
The reconstructions in Fig.~\ref{fig:recon} show intriguing deviations from $\Lambda$CDM that are allowed by the data. These reconstructions offer various plausible and non-exhaustive examples of the expansion history of the Universe allowed by the JLA data. The scatter in these plots do not reflect the posterior scatter and so $\Lambda$CDM is still in good agreement with the JLA data. As we mentioned earlier, we will perform some model selection analysis in future works.

\section{Summary and Discussion}\label{sec:dis}

We used the JLA SN Ia compilation to constrain the light-curve hyper-parameters for three different cosmological models, $\Lambda$CDM, CPL, and PEDE, and for the iterative smoothing method.  We find that the constraints on light-curve hyper-parameters for these models and methods are consistent with each other.
The fact these contours exactly overlap indicates the distance moduli of the JLA data might be independent of any cosmological assumptions.
Looking for the cause of this phenomenon will be a task for future study. 

We also reconstructed the distance modulus, expansion history of the Universe, $Om$ parameter, and deceleration parameter. These quantities are in good agreement with the predictions of the flat $\LCDM$ model, though the iterative smoothing method calculates example functions that deviate from $\Lambda$CDM yet are still allowed by the JLA data.

We can perform the same analysis for future supernova compilations, such as the ones from Large Synoptic Survey Telescope \citep{ivezic-etal19}, and the Wide Field Infrared Survey Telescope \citep{green-etal12,spergel-etal15}. These surveys may help us to detect any possible deviation from the standard supernova type Ia light-curve modeling as well as deviation from the standard $\LCDM$ cosmological model.

\acknowledgments

We thank Eric Linder for useful discussions.
This work was supported by the high performance computing clusters Seondeok at the Korea Astronomy and Space Science Institute. A.~S. would like to acknowledge the support of the Korea Institute for Advanced Study (KIAS) grant funded by the government of Korea. B.~L. would like to acknowledge the support of the National Research Foundation of Korea (NRF-2019R1I1A1A01063740).

\end{document}